\documentclass[preprint]{aastex6}
\usepackage{graphicx}
\usepackage{natbib}
\usepackage{amsmath}

\usepackage{color}
\definecolor{red}{rgb}{1.,0.0,0.}

\begin{document}                                                                     

\title{ Mass of Cepheid V350 Sgr Incorporating Interferometry and the Companion Mass
  \footnote{ }
} 


\author{Nancy Remage Evans}
\affil{Smithsonian Astrophysical Observatory,
MS 4, 60 Garden St., Cambridge, MA 02138; nevans@cfa.harvard.edu}

\author{Alexandre Gallenne}
\affiliation{Instituto de Alta Investigaci\'on, Universidad de Tarapac\'a, Casilla 7D, Arica, Chile}

\author{Pierre Kervella}
\affil{LIRA, Observatoire de Paris, Universit\'e PSL, CNRS, Sorbonne Universit\'e, Universit\'e de Paris, 5 Place Jules Janssen, 92195 Meudon, France
  and French-Chilean Laboratory for Astronomy, IRL 3386, CNRS and U. de Chile,
Casilla 36-D, Santiago, Chile}

\author{H. Moritz G\"unther}
\affil{Massachusetts Institute of Technology, Kavli Institute for Astrophysics and
Space Research, 77 Massachusetts Ave, NE83-569, Cambridge MA 02139, USA}

\author{Joanna Kuraszkiewicz}
\affil{Smithsonian Astrophysical Observatory,
MS 67, 60 Garden St., Cambridge, MA 02138; jkuraszkiewicz@cfa.harvard.edu}

\author{Richard I. Anderson}
 \affil{Institute of Physics, \'Ecole Polytechnique F\'ed\'erale de Lausanne (EPFL), Observatoire de Sauverny, Chemin Pegasi 51B, 1290 Versoix, Switzerland}
 
\author{Charles Proffitt}
\affiliation{Space Telescope Science Institute, 3700 San Martin Drive, Baltimore, MD 21218}

\author{Antoine M\'erand}
 \affil{European Southern Observatory, Karl-Schwarzschild-Str. 2, 85748 Garching, Germany}

\author{Boris Trahin}
\affil{LIRA, Observatoire de Paris, Universit\'e PSL, CNRS, Sorbonne Universit\'e, Universit\'e de Paris, 5 Place Jules Janssen, 92195 Meudon, France
and Space Telescope Science Institute, 3700 San Martin Drive, Baltimore, MD 21218}

\author{Giordano Viviani}   
\affil{Institute of Physics, \'Ecole Polytechnique F\'ed\'erale de Lausanne (EPFL), Observatoire de Sauverny, Chemin Pegasi 51B, 1290 Versoix, Switzerland}

\author{Shreeya Shetye}   
\affil{  Instituut
voor Sterrenkunde, KU Leuven, Celestijnenlaan 200D bus 2401, Leuven, 3001,
Belgium}


\begin{abstract}
The system V350 Sgr  has a classical Cepheid
for the primary.  Interferometry is presented for the system
and  the full orbit is determined.
The mass of the companion has
  been determined from an {\it IUE} spectrum and comparison with the
  mass-temperature relation from Detached Eclipsing Binaries.
  Combined with
  the mass of the companion (2.6 $\pm$ 0.2  M$_\odot$), the mass of
  the Cepheid is determined to be 4.7 $\pm$ 0.8 M$_\odot$.  For systems
  with less complete information, 
  Cepheid masses can be determined from a single-lined spectroscopic orbit,
  {\it Gaia} proper motion anomalies, and the mass of the companion from the
  ultraviolet.    
  Uncertainties resulting from different approaches to mass determination are
  discussed, and are expected to be 
  reduced after the {\it Gaia} DR4 release.  Temperatures for Morgan
  Keenan (MK) standard stars from the ultraviolet are also provided.  
   
\end{abstract}




\section{Introduction}

For more than 100 years the Leavitt (Period-Luminosity) Law for classical Cepheids
has been the mainstay in extragalactic distance determination.  At present it is
prominent in the ``Hubble tension'', the conflict between Hubble constant based on
Cepheids and supernovae and that from the $\Lambda$CDM model constrained by
the Planck satellite cosmic microwave
background observations  (Riess et al. 2022).

Quantitative understanding of Cepheids, however, has some challenges.
The first hydrodynamical pulsation calculations found a conflict
between masses derived from these calculations and those from evolutionary tracks
in the sense that evolutionary tracks predict higher masses than
pulsation calculations, as summarized  in Nielson et al. (2011).   Bono, et al. (2024)
provide a comprehensive summary of properties of stars in the
instability strip. The inclusion of rotation in evolutionary calculations
(Anderson, et al. 2014) improves the agreement between
evolutionary tracks and  measured masses.

In  recent years several problems and also tools have become prominent in
relation to this question.  Astroseismology provides insights into the internal
structure of Cepheid progenitors (B Stars)  including internal rotational
structure. This is relevant since overshoot from the B star convective core is
one way to bring better agreement between evolutionary tracks and pulsation masses.
Another complication is the detection of non-radial pulsation modes in addition
to the dominant radial pulsation mode (eg Netzel, Anderson, and Viviani 2024;
S\"uveges and Anderson 2018).

In addition, like many massive--intermediate mass stars,  Cepheids are found in multiple
star systems, often triple systems.  This can have several effects.  Since Cepheids are
post-red giant branch stars, a substantial fraction  are merger
products, with a number of others having undergone mass exchange (Sana, et al. 2012
for massive stars).
Furthermore  multiple systems may have had dynamical evolution of the orbital
parameters.  Several groups have modeled the distribution of binary components
(e.g. Frantisek, Anderson, and Kroupa 2024; Karczmarek et al. 2022).

Cepheids, on the other hand, have not lost a large fraction of their mass as more massive
O stars have.  If there has been mass loss, it is elusive. See, for example, the summary in
Matthews, et al. (2016).

A typical Cepheid has a mass of 5M$_\odot$.  However, the more luminous may have masses
up to 10M$_\odot$  (Evans, et al. 2013).  That is, most Cepheids will  become white
dwarfs, but they straddle the divide where the more massive may become core collapse
supernovae and neutron stars.  The frequency distribution of Cepheid periods (luminosities) is
well known, hence a mass calibration provides information of the relative frequency
of these objects.    

Measured masses are needed to investigate
these questions and to provide accurate predictions for the next
stages of evolution.  The list of Cepheids with orbits has now lengthened 
and orbits have been refined (Shetye et al. 2024).
Masses of Cepheids can be determined from favorable examples of these in two ways.
``Favorable'' means if they are reasonably bright and if the companion is reasonably massive
and hot.  High resolution spectra of the companions can be obtained using the Hubble
Space Telescope (HST).  This provides the mass ratio between the components from the
orbital velocity ratio. The Cepheid mass is derived by combining the mass ratio
with the mass of the companion from the spectral energy distribution.  The second approach
is resolving the system through interferometry to obtain the separation and the
inclination of the orbit.  Combining this with the distance provides the mass of the
Cepheid.  The current results are summarized in Gallenne, et al. (2025), including
SU Cyg, V1334 Cyg (Gallenne, et al. 2018b) and Polaris (Evans, et al. 2024a).
 These masses from  Milky Way Cepheids can be compared with masses from eclipsing binaries
 in the LMC (Pilecki et al. 2018).

The present paper is part of a series  to obtain Cepheid masses by combining spectroscopic
orbits of single lined binaries  with the masses of the binary secondaries.   The DR4 data release
of {\it Gaia} (Gaia collaboration 2016)  will provide inclinations for the
systems from proper motion anomalies,
completing the information
necessary to derive the mass of the Cepheid primary.  The discussion of the Cepheid FN Vel (Evans, et al.
2024b) sets out the first part of the method.  Masses of the companion are determined from ultraviolet
spectra where a hot companion dominates.  For FN Vel, a Space Telescope Imaging Spectrograph (STIS)
spectrum is used.  The temperature of
the companion is determined by matching it with BOSZ model atmospheres (Bohlin, et al. 2017).
A mass is inferred from this temperature using the mass-temperature relation for Detached
Eclipsing Binaries (DEBs) for B and early A stars (Evans, et al. 2023), in this case using
low resolution spectra from the {\it International Ultraviolet Explorer (IUE)} satellite.
During the analysis of the FN Vel system, additional BOSZ atmospheres were released (Meszaros, et
al. 2024), in which opacities and fluxes in the 1150 to 1900 \AA\/ region had been updated.
Because these model fluxes are evolving, in this series the approach is to use consistent
atmospheres (Bohlin et al. 2017) to make the comparison between the Cepheid companion
temperatures and the DEB temperatures to infer a mass.    Insuring consistency in
the chain of deductions is continued with the analysis in  Appendix A below, where
temperatures are determined for Morgan Keenan (MK) standard stars.  

V350 Sgr is among the Cepheids in binary systems suitable for mass determination.
Its binary motion was recognized by Lloyd Evans (1982) and Gieren (1982).
Subsequent velocities were combined in an orbital solution by Evans, et al. (2011)
and Gallenne, et al. (2019).  The system has also been successfully resolved with
the Very Large Telescope Interferometer (VLTI) 
 Precision Integrated Optics Near-infrared Imaging ExpeRiment
 (PIONIER) in H band (Gallenne, et al 2019).

In addition, velocities for the companion were measured in the ultraviolet, first
with the {\it HST} Goddard High Resolution Spectrograph 
(GHRS; Evans, et al. 1997) and subsequently with {\it HST} Space Telescope Imaging
Spectrograph (STIS) at higher resolution (Evans, et al 2018).
The data are available at MAST: \dataset[doi:10.17909/p0wr-x114].
The mass of the
companion V350 Sgr B is needed to combine with orbital velocity ratio of the Cepheid
and the companion to determine the Cepheid mass.  The temperature of the companion was
derived from comparison with {\it IUE} low resolution spectra (Evans and Sugars 1997) and
BOSZ atmospheres (Evans, et al. 2018).  The mass was determined from the mass--temperature
relation from Detached Eclipsing Binaries (DEBs; Torres, Andersen, and Gimenez 
2010).  The most recent result for the Cepheid mass is 5.2 $\pm$ 0.3  M$_\odot$
(Evans, et al. 2018).

The current study presents mass determination for V350 Sgr, a system 
which has one of the most complete suites of observations for a system containing a Cepheid.
In addition to the spectroscopic orbit and the orbital velocity amplitude of the
companion from ultraviolet spectra, interferometry is presented, and also the
mass of the companion from a spectrum in the ultraviolet.  From this  the mass
of the Cepheid is derived.
The V350 Sgr system   has a long enough orbital period (4.0 years) that the inclination
derived by Kervella, et al. (2019) from the proper motion anomaly from {\it Gaia} DR2
( {\it Gaia} Collaboration et al. (2018))
and {\it Hipparcos} are reasonably accurate.  An important input is the mass of the
secondary V350 Sgr B.  The present discussion updates this in two ways.  First, the
temperature is determined with the same software  using BOSZ atmospheres as
was used for FN Vel (Evans et al, 2024b).  Second, the mass-temperature
relation was based on {\it IUE} spectra of DEBs.   One aim of this study is to compare
the Cepheid mass derived from {\it Gaia} and a companion mass with the Cepheid mass from
other approaches.  This approach provides a longer list of Cepheids systems for which
masses can be determined.  

Sections below discuss the interferometry, the determination of the mass of the
companion, the orbit, and the results.  
In addition, temperatures from {\it IUE}
spectra of  stars which define the Morgan Keenan (MK) spectral classes are derived.  These
temperatures then relate the temperatures of Cepheid companions directly to
calibrations such as the luminosity calibration.






	\section{Interferometric observations and data reduction}
	\subsection{Data acquisition}
	
	We collected long-baseline interferometric data with the four-telescope combiner PIONIER
        (Le Bouquin, et al. 2011) installed at the VLTI (Haubois et al. 2022). PIONIER combines the light coming from four telescopes in the $H$ band, either in a broad band mode or with a low spectral resolution, where the light is dispersed across several spectral channels. The recombination provides simultaneously six visibilities and four closure phase signals per spectral channel.
	
	Our observations were carried out from 2013 to 2024 using the 1.8 m Auxiliary Telescopes with the largest available configurations, providing six projected baselines ranging from 40 to 200\,m. We used the \emph{GRISM} mode where the fringes are dispersed into six spectral channels (three before December 2014). We monitored the interferometric transfer function with the standard procedure which consists of interleaving the science target by reference stars. The calibrators were selected using the SearchCal software\footnote{\url{http://www.jmmc.fr/searchcal}} (Bonneau, et al., 2011) provided by the Jean-Marie Mariotti Center, and are listed in Table~\ref{table__log}, together with the journal of the observations.

	\begin{deluxetable*}{cccccc}
		\tablecaption{\label{table__log} Log of our PIONIER observations.}
		\tablewidth{0pt}
		\tablehead{
		\colhead{UT Date}  & \colhead{JD} & \colhead{$N_\mathrm{spec}$} & \colhead{$N_\mathrm{vis}$} & \colhead{$N_\mathrm{CP}$} & \colhead{Calibrators}  }
		\startdata
		2013 Jul. 11   &  2456484.74659   &  3 &  630 &  396 &  1,2 \\
		2013 Jul. 15  &  2456488.62701    &  3 &  828 & 546  &  1,3  \\
		2019 May 20 &  2458623.89263  & 6  & 679  &  550  &  4,5,6 \\
		2021 Aug. 17 &  2459443.61811    & 6 & 1440 & 960 &  4,5,6 \\
		2024 Jul. 15 & 2460506.65775    & 6 & 1911 & 1308 & {\bf 7, 8, 9} \\
		\enddata
		\tablecomments{$N_\mathrm{spec}$: number of spectral channel. $N_\mathrm{vis}$: number of visibility measurements. $N_\mathrm{CP}$: number of closure phase measurements. The calibrators used have the following angular diameters: 
			1: $\theta\mathrm{_{LD}(HD174774)} = 1.103\pm0.015$\,mas, 2: $\theta\mathrm{_{LD}(HD172051)} = 0.637\pm0.045$\,mas, 3: $\theta\mathrm{_{LD}(HD171960)} = 1.121\pm0.016$\,mas, 4: $\theta\mathrm{_{LD}(HD173228)} = 0.382\pm0.009$\,mas, 5: $\theta\mathrm{_{LD}(HD174423)} = 0.405\pm0.010$\,mas, 6: $\theta\mathrm{_{LD}(HD174560)} = 0.392\pm0.009$\,mas, 7: $\theta\mathrm{_{LD}(HD174134)} = 0.354\pm0.008$\,mas, 8: $\theta\mathrm{_{LD}(HD172034)} = 0.360\pm0.008$\,mas, 9: $\theta\mathrm{_{LD}(HD172545)} = 0.352\pm0.000$\,mas.
		}
	\end{deluxetable*}
	
\begin{figure*}[ht]
		\centering
		\resizebox{\hsize}{!}{\includegraphics{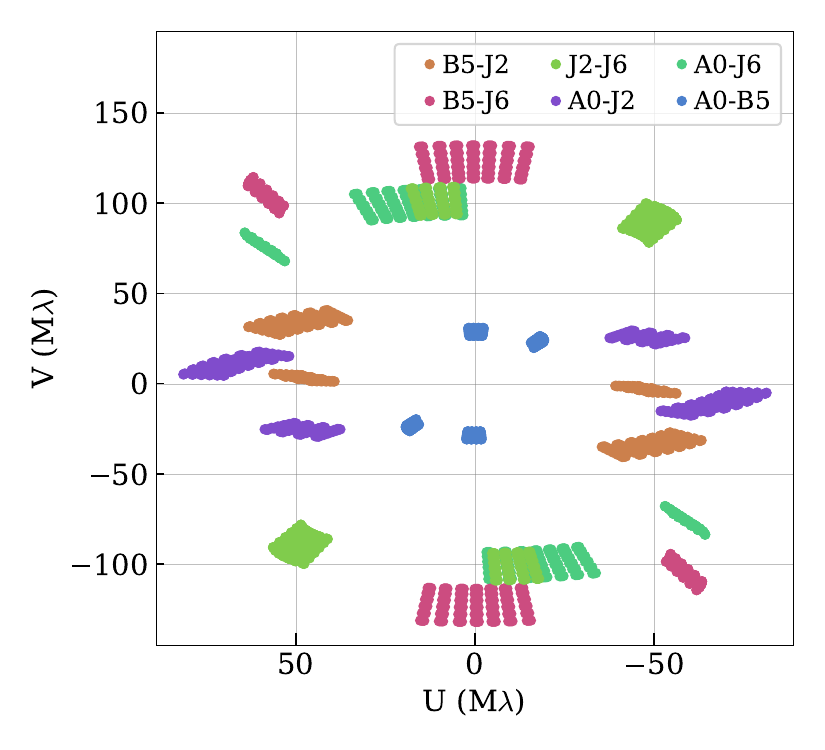}\includegraphics{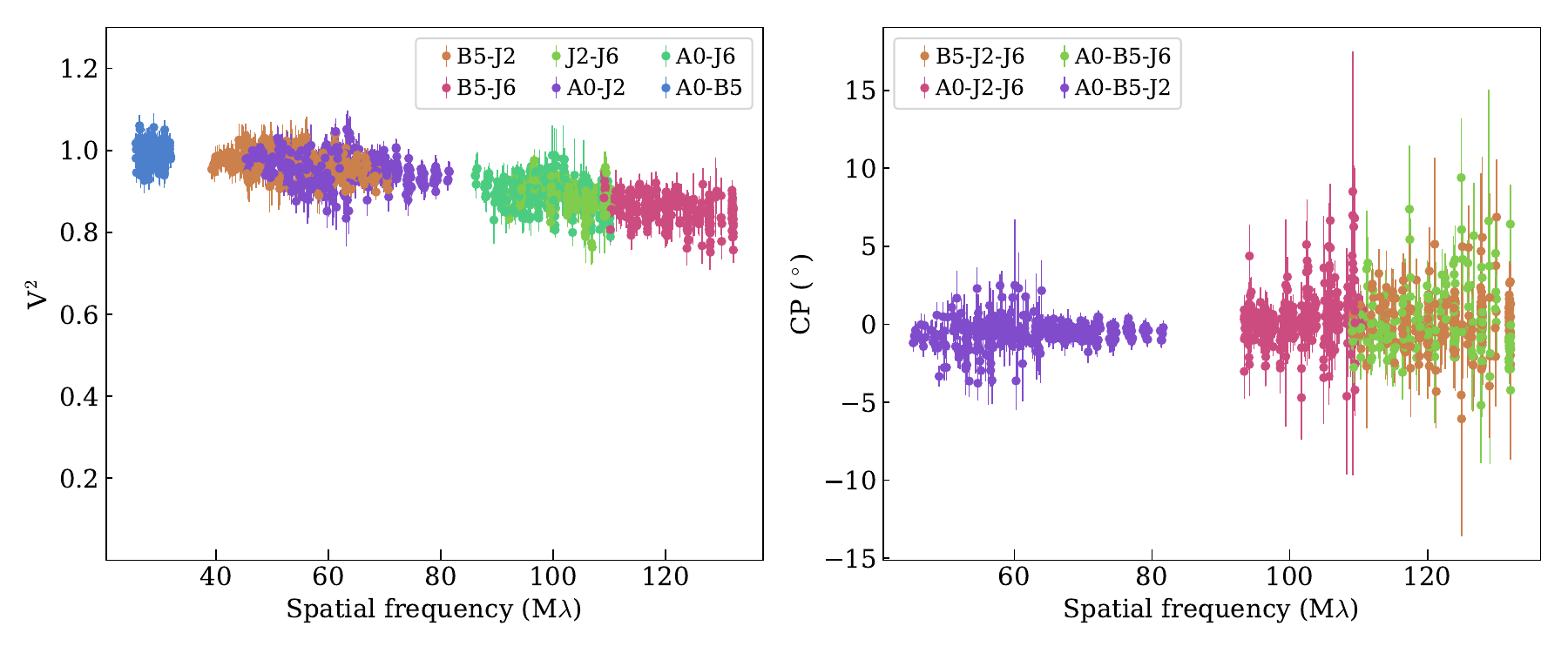}}
		\caption{$(u,v)$ coverage, squared visibilities and closure phase data for the observations of 2024 July 15.}
		\label{figure__data}
	\end{figure*}

	The data have been reduced with the \emph{pndrs} package described in Le Bouquin, et al. (2011). The main procedure is to compute squared visibilities and triple products for each baseline and spectral channel, and to correct for photon and readout noises. The last data set acquired is displayed in Fig.~\ref{figure__data}.
	
	\subsection{Data analysis}
	
	To detect the companion, we used the interferometric tool CANDID\footnote{Available on Github, \href{https://doi.org/10.5281/zenodo.15334366}{doi:10.5281/zenodo.15334367}, and \href{https://doi.org/10.5281/zenodo.15334372}{doi:10.5281/zenodo.15334373} for a GUI version.}
        (Gallenne, et al. 2015). The main function allows a systematic search for companions performing an $N \times N$ grid of fits, the minimum required grid resolution of which is estimated a posteriori in order to find the global minimum in $\chi^2$. The tool delivers the binary parameters, namely the flux ratio $f$, and the relative astrometric separation $(\Delta \alpha, \Delta \delta)$, together with the uniform-disk angular diameter $\theta_\mathrm{UD}$ of the primary star (the Cepheid). The significance of the detection is also given, taking into account the reduced $\chi^2$ and the number of degrees of freedom. They are listed in Table~\ref{table__astrometry_results}, together with our measured astrometric positions. 
	
	For the observations of 2013, we combined the dataset in order to increase the detection level (although the companion is detected for each epoch, but suffered of large astrometric uncertainties). In this case, we only used the closure phase because of the variation of the angular diameter between these two observations (which was fixed to the average fitted value between the two epochs).
	
	Uncertainties of the fitted parameters are estimated using a bootstrapping function. From the distribution, we took the median value and the maximum value between the 16th and 84th percentiles as uncertainty for the flux ratio and angular diameter. For the fitted astrometric position, the error ellipse is derived from the bootstrap sample (using a principal components analysis). We quadratically added a conservative systematic error of $\delta \lambda = 0.35$\,\% to the astrometry, as estimated by Gallenne, et al. (2018a), to take into account the uncertainty from the wavelength calibration. Uncertainty of the angular diameter measurements was estimated using the conservative formalism of Boffin, et al. (2014) as follows:
	\begin{displaymath}
		\sigma^2_\mathrm{\theta_{UD}} = N_\mathrm{sp} \sigma^2_\mathrm{stat} + \delta \lambda^2 \theta^2_\mathrm{\theta_{UD}}
	\end{displaymath}
	where $N_{sp}$ is the number of spectral channels and $\sigma^2_\mathrm{stat}$ the uncertainty from the bootstrapping. Measurements are also listed in Table~\ref{table__astrometry_results}.
The columns in Table~\ref{table__astrometry_results} are the date, the relative astrometric position of the companion $\Delta \alpha$ and $\Delta \delta$, the parameters of the error ellipse, that are the semi-major axis $\sigma_\mathrm{maj}$, the semi-minor axis $\sigma_\mathrm{min}$, and the position angle $\sigma_\mathrm{PA}$ measured from north to east, the flux ratio in $H$, the uniform disk angular diameter, and the detection level of the companion in number of sigma.         
	
	We measured a mean uniform disk diameter in the $H$ band of $\mathrm{\theta_{UD}} = 0.371\pm0.029$\,mas (the standard deviation is taken as uncertainty), which is consistent with the $0.42$\,mas estimated by Trahin, et al. (2021) from a spectro-photometric analysis. We also estimated an average flux ratio in $H$ of $f_\mathrm{H} = 0.51\pm0.05$\,\%.

		\begin{deluxetable*}{ccccccccc}

		\tablecaption{Relative astrometric position of the V350~Sgr companion.\label{table__astrometry_results}.}
		\tablewidth{0pt}
		\tablehead{
			\colhead{HJD} & \colhead{$\Delta \alpha$}	&  \colhead{$\Delta \delta$} & \colhead{$\sigma_\mathrm{PA}$}	& \colhead{$\sigma_\mathrm{maj}$}	&  \colhead{$\sigma_\mathrm{min}$}	&  \colhead{$f$} &  \colhead{$\theta_\mathrm{UD}$} & \colhead{$n\sigma$}	\\
			\colhead{(Day)} & \colhead{(mas)} & \colhead{(mas)} & \colhead{(deg)} & \colhead{(mas)} & \colhead{(mas)} & \colhead{($\%$)}  & \colhead{(mas)} & 
		}
		\startdata
		2456486.9765 & 0.8580 & 2.7861 & 123.9 & 0.2192 & 0.0862   & $0.52\pm0.06$  & 0.49 & 10  \\
		2458623.8926 & 1.3798 & $-4.4549$ & 94.1 & 0.2041 & 0.1090  & $0.58\pm0.10$ & $0.344\pm0.500$ & 6.3 \\
		2459443.6224 & 1.5764 & 2.9278 & 62.2 & 0.1116 & 0.0524                       & $0.51\pm0.07$  & $0.357\pm0.028$ & 9.0 \\
		2460506.6635 & $-3.2332$ & $-4.1234$ & $-173.7$ & 0.1499 & 0.0638  & $0.44\pm0.06$  & $0.411\pm0.018$ & 4.6 \\
		\enddata
		\end{deluxetable*}

\section{The Companion V350 Sgr B}                

\subsection{The Observation}

The temperature of the companion V350 Sgr B was determined using the {\it IUE} spectrum
SWP 44358 from 1150 to 1900 \AA.  It has been previously discussed by Evans and Sugars
(1997) and Evans et al. (2018).  It was reduced in the same way as the spectra for
the DEB mass-temperature relation (Evans, et al. 2023).  This includes correction of the
fluxes to the {\it HST} STIS calibration, removing (``blemishing'') geocoronal
Ly $\alpha$ flux and applying a 10 point boxcar smooth.  



\subsection{Energy Distributions}

\subsubsection{E(B-V)}

The first step in analyzing the energy distribution of V350 Sgr B
 is to determine the reddening of the system.  For a reasonably bright
companion this starts with assessing the contribution to the colors from the
companion, as summarized in Table~\ref{corr}.  Observed composite magnitudes are taken from
Berdnikov, et al, 2000.  The procedure is to start with E(B-V) from the
uncorrected colors (since the contribution of the companion is modest)
using the relation from Fernie (1990).
The  temperature is recalculated from the unreddened {\it IUE} spectrum.  New
corrected colors are then used to confirm that the E(B-V) is accurate.
The colors and reddening are provided in Tables~\ref{corr} and ~\ref{ebv}.
The temperature from the preliminary fit is 11500 K $\pm$ 500 K (uncertainty from
visual assessment). This corresponds to B9 V using the temperatures from
the MK standards in Table~\ref{tmk}.  
From the calibration of Drilling and Landolt (2000), this corresponds to
M$_V$ = +0.20 mag.  For the Cepheid, the calibration of the Leavitt Law from 
Cruz Reyes and Anderson (2023) is M$_V$ = -3.84.
Using this magnitude difference, the original reddening, and the colors
of a B9 V star, the magnitudes for the Cepheid alone are found (Table~\ref{corr}),
and from this the revised reddening in Table~\ref{ebv}.

\begin{deluxetable}{lrrr}
\tablecaption{Correction for the Companion\label{corr}}
\tablewidth{0pt}
\tablehead{
  \colhead{} &  \colhead{ $<$B$>$ }  & \colhead{$<$V$>$}  & \colhead{$<$I$>$} \\
  \colhead{} &  \colhead{ mag }  & \colhead{mag}  & \colhead{mag} \\
}
\startdata
Cepheid + Companion  &  8.382   &   7.474  &    6.422  \\  
Cepheid               & 8.46    &   7.51   &    6.45  \\
Companion            &  11.36   &   11.13  &   10.76  \\
\enddata
\end{deluxetable}

\begin{deluxetable}{lrrr}
\tablecaption{Reddening\label{ebv}}
\tablewidth{0pt}
\tablehead{
  \colhead{} &  \colhead{  $<$B$>$ - $<$V$>$ }  & \colhead{ $<$V$>$ - $<$I$>$}  & \colhead{E(B-V) } \\
  \colhead{} &  \colhead{ mag }  & \colhead{mag}  & \colhead{mag} \\
}
\startdata
Original   &      0.91  &            1.05      &      0.30  \\
Corrected   &     0.95   &            1.06     &        0.28  \\
\enddata
\end{deluxetable}

\subsubsection{Companion Temperature}

The companion temperature was determined in the same way as for FN Vel
(Evans, et al. 2024b).  To maintain consistency with
the mass--temperature calibration from DEBs (Evans, et al. 2023),
BOSZ atmospheres (Bohlin, et al. 2017) are used.  To model
late B and early A stars (typical of Cepheid companions) the
atmospheres selected have solar abundance, surface gravity
(log g) of 4.0, microturbulence of 2 km s$^{-1}$, and
instrumental broadening of 500 km s$^{-1}$.  

\begin{figure}
  \plotone{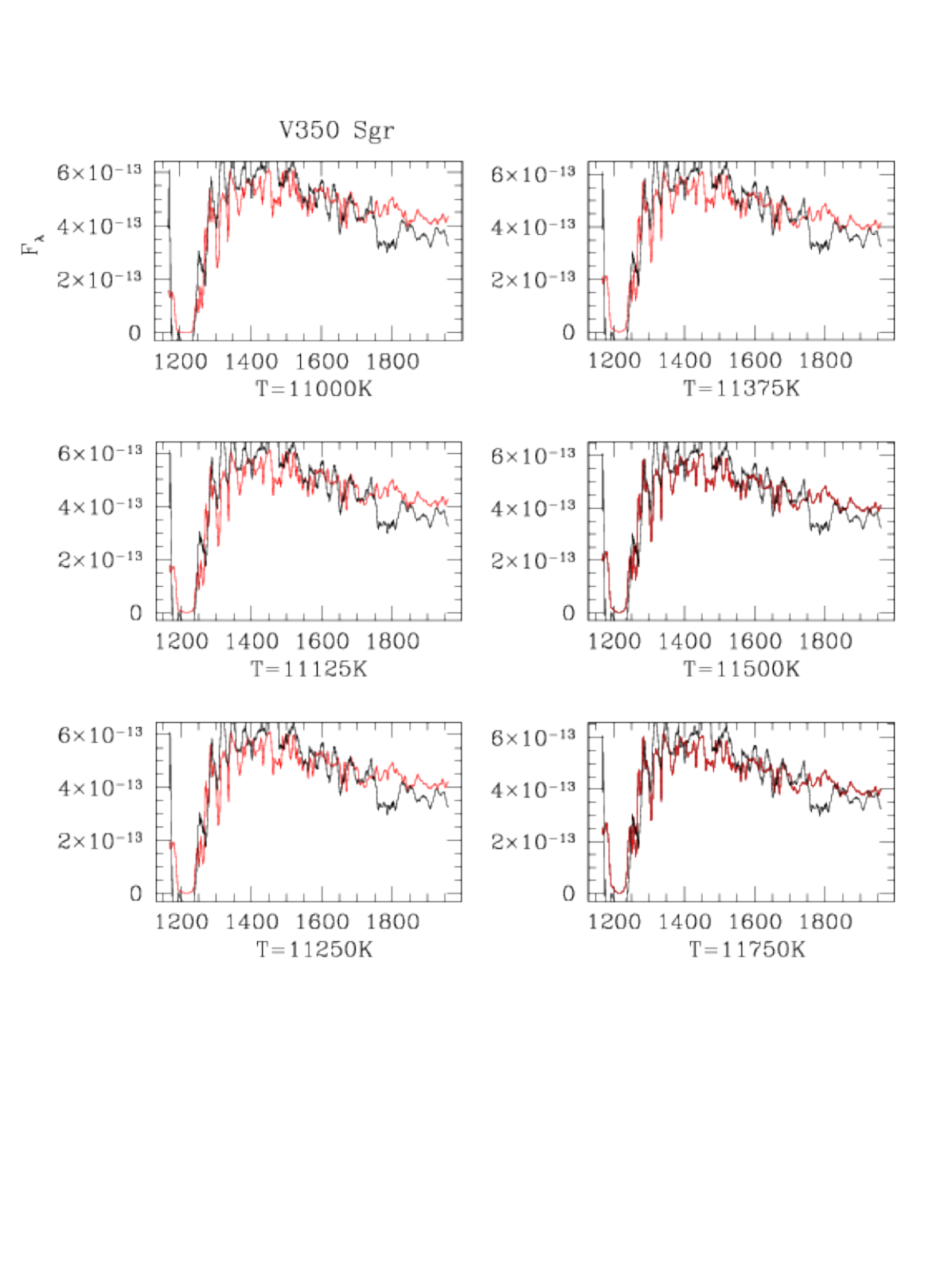}
\caption{Energy distribution comparisons for V350 Sgr.  Comparisons are for the
    spectrum (black) with model atmospheres (red) for a series of temperatures from
        11000 to 11750 K. The X axis is wavelength in \AA; 
  the  Y axis is flux in ergs cm$^{-2}$ s$^{-1}$ \AA$^{-1}$.
    \label{v350.sgr}}
\end{figure}

\begin{figure}
  \plottwo{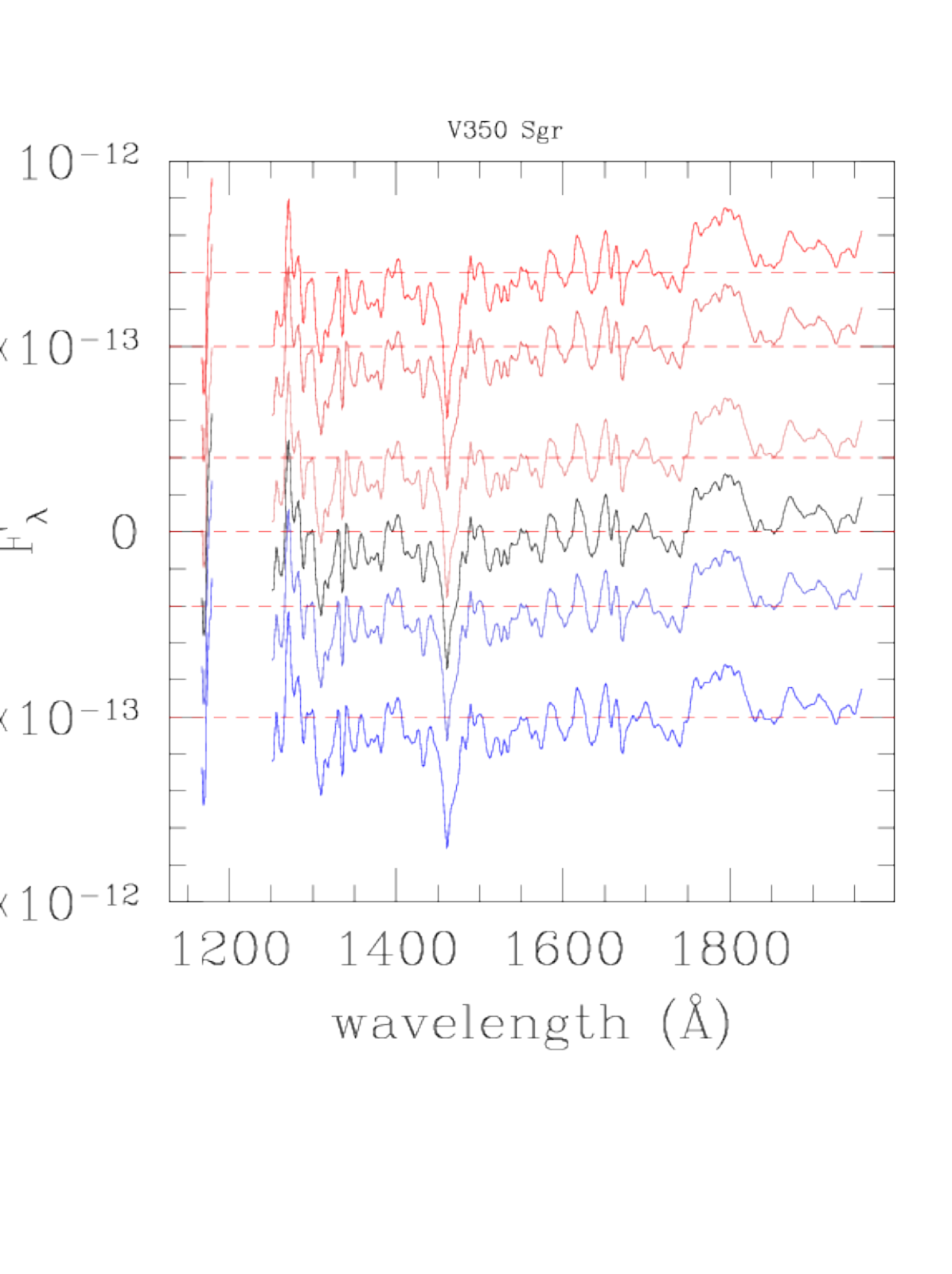}{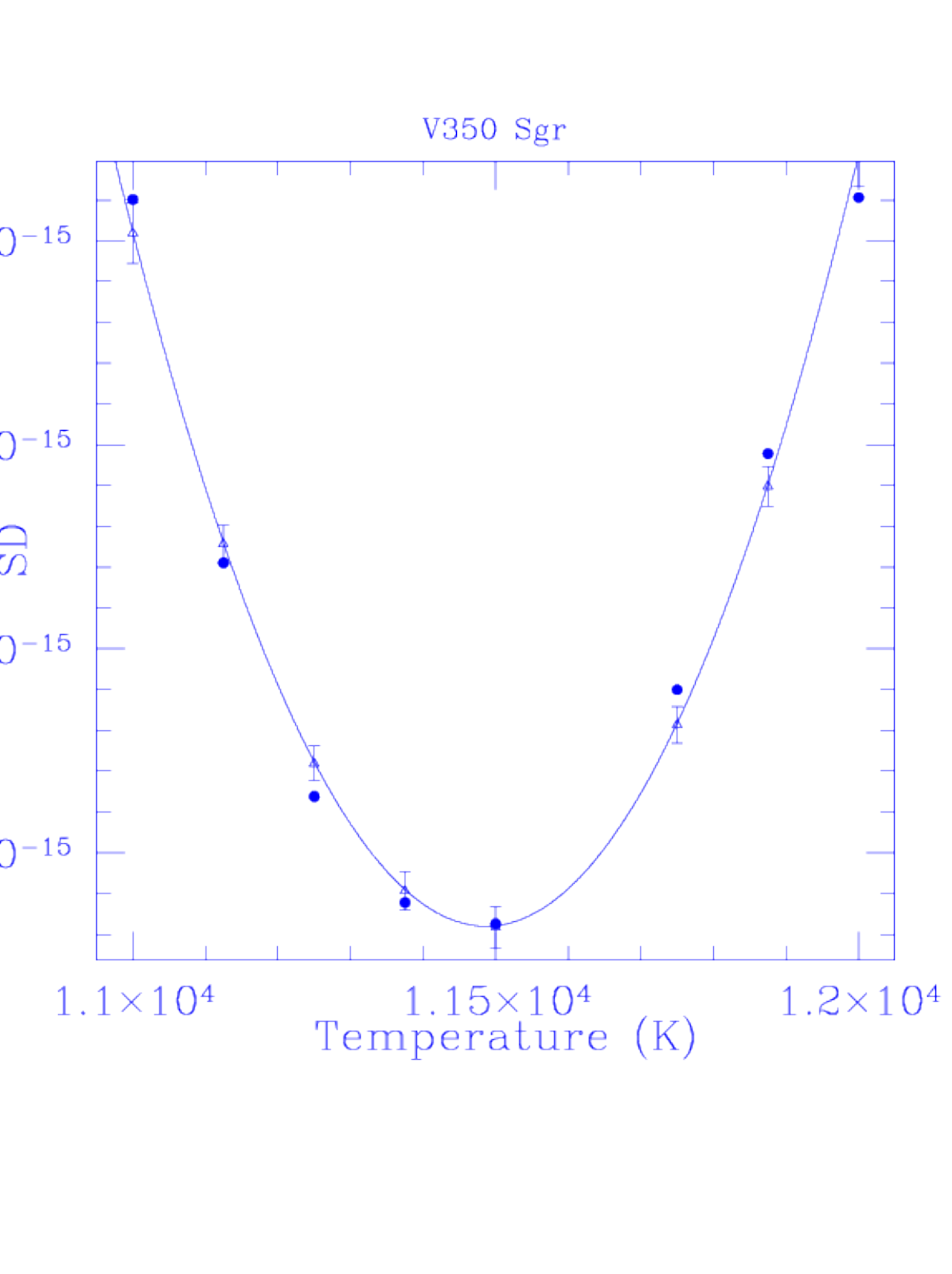}
  \caption{Energy distribution  V350 Sgr.  Left:  The difference between the
   model  and the spectrum.  Model temperatures (starting from the top) are 11000 11125
    11250 11375 11500 11750  K. The X axis is wavelength in \AA; 
    the  Y axis is flux difference in ergs cm$^{-2}$ s$^{-1}$ \AA$^{-1}$;
    the models are offset for clarity. 
 Right: The standard deviations from the spectrum-model comparison for
  V350 Sgr as the
  temperature of the models is changed.  Dots: the values of standard deviation;
  triangles: the parabola fit.
\label{v350sgr}}
\end{figure}

Table~\ref{ebv} confirms the reddening based on the preliminary temperature.
The temperature from the solution is thus 11500 K $\pm$ 500 K.
Fig~\ref{v350.sgr} shows the temperature comparisons for a series of
model temperatures.  Fig~\ref{v350sgr} shows the flux differences between
them as a function of wavelength, and also the parabola fit to the
standard deviation between the spectrum  and the model. 

\hfill\vfill
\eject

To further illustrate the temperature determination approach
Appendix A shows the results for ultraviolet spectra of
MK spectral classes.  This allows comparison with calibrations
from other wavelength regions.

\subsubsection{Companion Mass}

Using this temperature, the mass of the companion was estimated  in the
same way as for FN Vel (Evans, et al. 2024b) using the mass--temperature
relation for DEBs (Evans, et al. 2023).  For Cepheid companions there is
extra information about age, which provides the constraint that they must
be in the younger half of the DEB mass--temperature sample.  For this
reason a mass--temperature relation which is  0.02 smaller in log M is
used.  The range of masses for $\pm$ 500 K is shown in Table~\ref{m2}.  
The scatter around the  mass--temperature
relation is estimated as for FN Vel
to be half the rms scatter for the DEB sample because of the age
constraint, which is 0.02 in log M.  This is less than the uncertainty in
the temperature, hence the total estimated uncertainty is 0.035 in log M.
From Table~\ref{m2} the companion mass is 2.6 M$_\odot$ $\pm$ 7\%.

\begin{deluxetable}{lr}
\tablecaption{The Mass of V350 Sgr B\label{m2}}
\tablewidth{0pt}
\tablehead{
  \colhead{Temperature} &  \colhead{Mass  } \\  
  \colhead{ K} &  \colhead{ M$_\odot$  } \\
}
\startdata
11000  &  2.42  \\
11500  &  2.60  \\
12000  &  2.77  \\
\enddata
\end{deluxetable}

\section{Orbit}
\label{section__model_fitting}

Orbital parameters are determined by simultaneously fitting the radial velocities (RVs) of the primary star (the Cepheid, which is a single-line spectroscopic binary) and the astrometric positions using a Markov chain Monte Carlo routine (MCMC, 100\,000 samples with uniform priors) \footnote{With the Python package \emph{emcee} developed by Foreman-Mackey, et al. (2013).}. Our RV models include the pulsation of the Cepheid and its orbital reflex motion due to the presence of a companion, while the astrometric model defines the relative astrometric motion of the companion around the Cepheid. Our fitting procedure is detailed in our previous works [see e.g. Gallenne, et al. (2025), Gallenne, et al. (2018b), Evans, et al. (2024c),
  Evans, et al. (2024a)].

\begin{figure*}[ht]
		\centering
		\resizebox{\hsize}{!}{\includegraphics{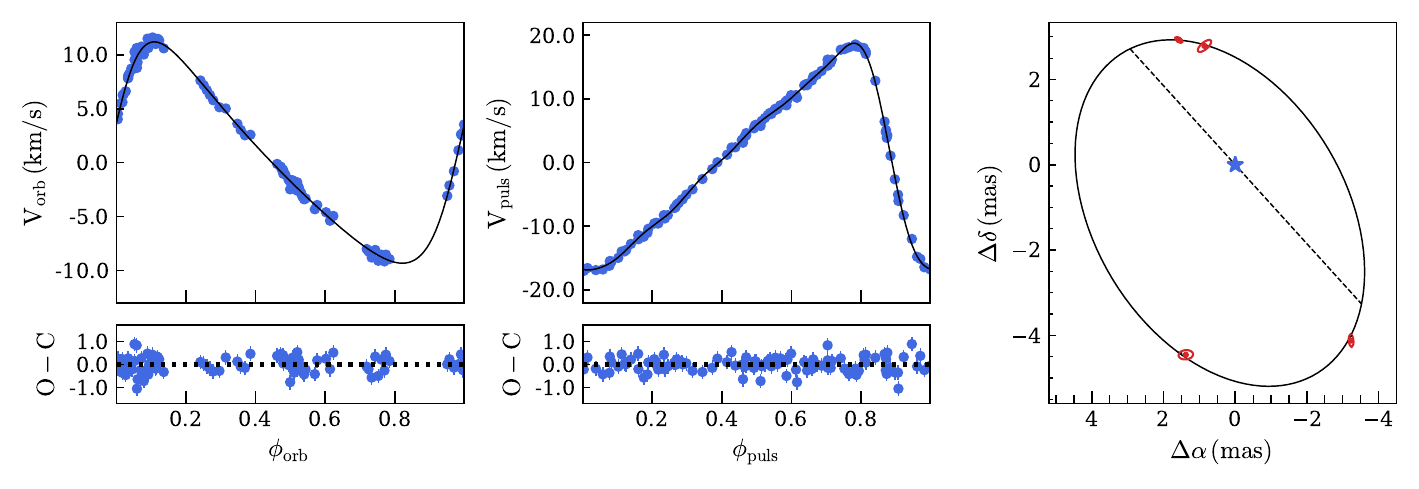}}
		\caption{Result of our combined fit. Left: fitted (solid lines) and measured (blue dots) Cepheid's orbital velocity. Middle: fitted and measured pulsation velocity. Right: relative astrometric orbit of V350~Sgr AB from VLTI.}
		\label{figure__orbit}
	\end{figure*}

	We used the RVs from Evans, et al. (2011), but only measurements from Eaton in that source which offer a good precision and orbital coverage. Because the orbital period is very close to 4 years, particular care must
        be taken to get good phase coverage.  
        We did not include additional data from the literature, as doing so would have increased the residual scatter due to potential systematic errors caused by zero-point offsets when combining different instruments. Our orbital fit is displayed in Fig.~\ref{figure__orbit} and the final parameters are listed in Table~\ref{table__results}. Spectroscopic parameters are consistent with previous works, as well as the orbital inclination with the value of Kervella, et al. (2019) from the proper motion anomalies.   We have also done an MCMC solution using the interferometry only. The parameters dependent on the interferometry (a, i, and $\Omega$) all have values and errors consistent with the values in Table~\ref{table__results}. 
	
	\begin{table}[!h]
		\centering
		\caption{Final estimated parameters of the V350~Sgr system. Index 1 designates the Cepheid and index 2 its companion. The A and B values are the Fourier coefficients of the radial velocity pulsation curve.}
		\begin{tabular}{cc|cc} 
			\hline
			\hline
			\multicolumn{2}{c|}{Pulsation} & \multicolumn{2}{c}{Orbit} \\
			\hline
			$P_\mathrm{puls}$ (days)	     &   $5.15424\pm0.00003$ 		&   $P_\mathrm{orb}$ (days)								& $1466.15\pm0.64$ 		  \\
			$T_0$ (JD)						   			  &   2440146.6156\tablenotemark{a}	&   $T_\mathrm{p}$ (JD)										& $2450556.17\pm0.59$   \\
			$A_1$ (km~s$^{-1}$)					&	 $-9.23\pm0.95$							&   $e$																     	&   $0.351\pm0.008$	\\
			$B_1$ (km~s$^{-1}$)					&	 $-11.78\pm0.77$						   &   $\omega$	($^\circ$)								  		&	$285.02\pm0.42$		  \\
			$A_2$ (km~s$^{-1}$)					&	 $-5.57\pm0.30$							&   $K_1$ ($\mathrm{km~s^{-1}}$)				   &	$10.269\pm0.015$	\\
			$B_2$ (km~s$^{-1}$)					&	 $-1.42\pm0.91$							&   $v_\gamma$	($\mathrm{km~s^{-1}}$)	  &	$11.476\pm0.022$		\\
			$A_3$ (km~s$^{-1}$)					&	 $-2.46\pm0.43$							 &   $\Omega$	($^\circ$)									 &	$47.2\pm1.3$	\\
			$B_3$ (km~s$^{-1}$)					&	 $1.51\pm0.62$						   &   $i$ ($^\circ$)													&	$48.34\pm3.05$	\\
			$A_4$ (km~s$^{-1}$)					&	 $-0.23\pm0.41$							&   $a$ (mas)														  &	$4.98\pm0.12$	\\
			$B_4$ (km~s$^{-1}$)					&	 $1.28\pm0.15$							&   $a_1 \sin i$ (au)												&	$1.296\pm0.004$	 \\
			$A_5$ (km~s$^{-1}$)					&	 $0.38\pm0.17$							&   $f(M_2)$ ($M_\odot$)  									&   $0.135\pm0.001$	\\
			$B_5$ (km~s$^{-1}$)					&	 $0.38\pm0.17$							&   &   \\
			$A_6$ (km~s$^{-1}$)					&	 $0.38\pm0.07$							&   &	\\
			$B_6$ (km~s$^{-1}$)					&	 $0.03\pm0.17$							&   &	\\
			\hline															
		\end{tabular}
		\tablenotetext{a}{Kept fixed to the value given in Evans, et al. (2011).}
		\label{table__results}
	\end{table}

	\section{Discussion}\label{discuss}
	
	V350~Sgr is a single-line spectroscopic binary, therefore the masses and the distance are degenerate parameters even when the inclination is known from the interferometry.
        Some additional information is needed to derive both masses.  One possibility  to resolve
        the degeneracy is to  assume either mass or distance in order to determine the other.  For the
        V350 Sgr system, there are several approaches possible.

      Information about the mass ratio between M$_1$ and  M$_2$ was provided by Evans, et al. (2018).  They obtained two  velocity measurements of the companion  from STIS high resolution spectra, which can be incorporated into our global fit. This allows us to independently estimate the system's masses and distance. This analysis is similar to previous works  (Gallenne, et al. (2025), Evans, et al. (2024a), Gallenne, et al. (2018b)), although it remains preliminary due to the limited number of measurements available. In this section we explore various methods to derive the masses and the distance.
	
	\subsection{Cepheid mass from the estimated companion mass}\label{m2.orb}
	
	From the orbital inclination and the companion mass determined in the previous Sections, the mass of the Cepheid can be estimated via the following equation:
	\begin{equation}
		M_1 = \sqrt{ \dfrac{2\pi G M_2^3 \sin^3 i}{P_\mathrm{orb} K_1^3 (1 - e^2)^{3/2}} } - M_2
	\end{equation}
	
	We generated 100~000 realizations for each measured quantity by sampling from a Gaussian distribution, using their respective values and uncertainties as the mean and standard deviation of the distribution. We obtained $M_1 = 4.7\pm0.8\,$M$_\odot$.

        One goal of this paper is to compare the mass determined for V350 Sgr from this orbit with a directly
        determined inclination with a Cepheid mass from the combination of an inclination from {\it Gaia}, the
        companion mass, and the single-lined spectrosopic orbit.  To most closely resemble other systems, we
        have used the spectroscopic orbit from Evans, et al. (2011), the inclination from Kervella, et al. (2019;
        i = 35 $\pm$ 12 $^o$), and the companion mass from above.  
        The Cepheid mass was 2.2 $\pm$ 2.0 M$_\odot$.
        A similar solution using the
        inclination 64 $\pm$ 9$^o$ from the proper motion anomaly from {\it Gaia} DR3 ({\it Gaia} Collaboration 2023)
        is 6.7 $\pm$ 1.3  M$_\odot$.
        In both cases the error is much larger than that from the orbit above, and
        they are only marginally in agreement with
        that mass.  The inclination from   {\it Gaia} DR3 is less accurate than that from DR2 since the
        orbital period (4 years) is very close to the DR3 window of 3 years, hence the smearing is very
        strong.  The value from DR2 is better because the {\it Gaia} averaging is on a shorter time window.
        The same analysis will be performed with {\it Gaia} DR4 results.  The V350 Sgr system has        a high Renormalized Unit Weight Error
          (RUWE) in DR3, hence it is likely that it will have a non-single star solution in DR4, which will provide an accurate parallax.  Combining  DR4 epoch astrometry with the existing interferometric and spectroscopic measurements will provide further improvement, which is  difficult to predict at this time. An example of a simulation for a high contrast system between DR3 and DR4 is provided in Table 3 in Wallace, et al. (2025).

	\subsection{Distance determination from estimated masses}
	
	Knowing the mass of each component and the semi-major axis, we can determine the Cepheid distance using  Kepler's third law. From the previous Gaussian distribution we find $d = 986\pm48$\,pc. This is $\sim 3\sigma$ smaller that the Gaia estimate discussed below.
	
	\subsection{Mass determination from assumed distance}
	
	We can first assume the distance and derive the individual masses. We took the parallax measured by
        {\it Gaia}, $\varpi = 0.810\pm0.062$\,mas ($1235\pm114$\,pc, including the $-0.036$\,mas correction from
        Lindegren, et al. 2021).  The RUWE of Gaia is 2.4,
         indicating a relatively poor astrometric solution
        because the system is a resolved binary. 
       We implemented the distance in our MCMC analysis with a normal distribution centered on 0.810\,mas with a
        standard deviation of 0.062\,mas. Masses are then derived from the distribution with:
		\begin{eqnarray*}
		M_\mathrm{T} &=& \dfrac{a^3}{P_\mathrm{orb}^2\,  \varpi^3} , \\
		q &=& \left[ \dfrac{a \sin{i}}{0.03357\, \varpi\, K_1\, P_\mathrm{orb} \sqrt{1 - e^2}} - 1 \right]^{-1} \\
		K_2 &=& \dfrac{K_1}{q} \\
		M_1 &=& \dfrac{M_\mathrm{T}}{1+q}\\
		M_2 &=& q\,M_1
	\end{eqnarray*}
	with $M_\mathrm{T} = M_1 + M_2$ the total mass in $M_\odot$, $a$ and $\varpi$ in mas, $P_\mathrm{orb}$ in years, $K_1$ in km\,s$^{-1}$ and $q = M_2/M_1$ the mass ratio. However, we find a mass for the Cepheid of $M_1 \sim 10\,M_\odot$, which is unexpectedly large for a 5 day Cepheid, and a secondary mass of $M_2 \sim 4\,M_\odot$ which is also not consistent with the mass estimate from the IUE spectra. This suggests that the distance given by {\it Gaia} is overestimated.
	
	We performed the same analysis using the distance $d = 940 \pm 100$\,pc from the period-luminosity relation of Gieren, et al.  (2018). We found a more consistent secondary mass of $\sim2.4\,M_\odot$ and a Cepheid mass of $\sim 4\,M_\odot$, which seems more consistent with what we expected for such a star.
        The result is similar to that found using the absolute magnitude from Cruz Reyes and
        Anderson (2023).  In summary, determinations using the system distance should be re-evaluated with the
        results from {\it Gaia} DR4 when available.

	\subsection{Mass and distance measurements from STIS spectra}
	
\begin{figure*}[ht]
		\centering
		\resizebox{\hsize}{!}{\includegraphics{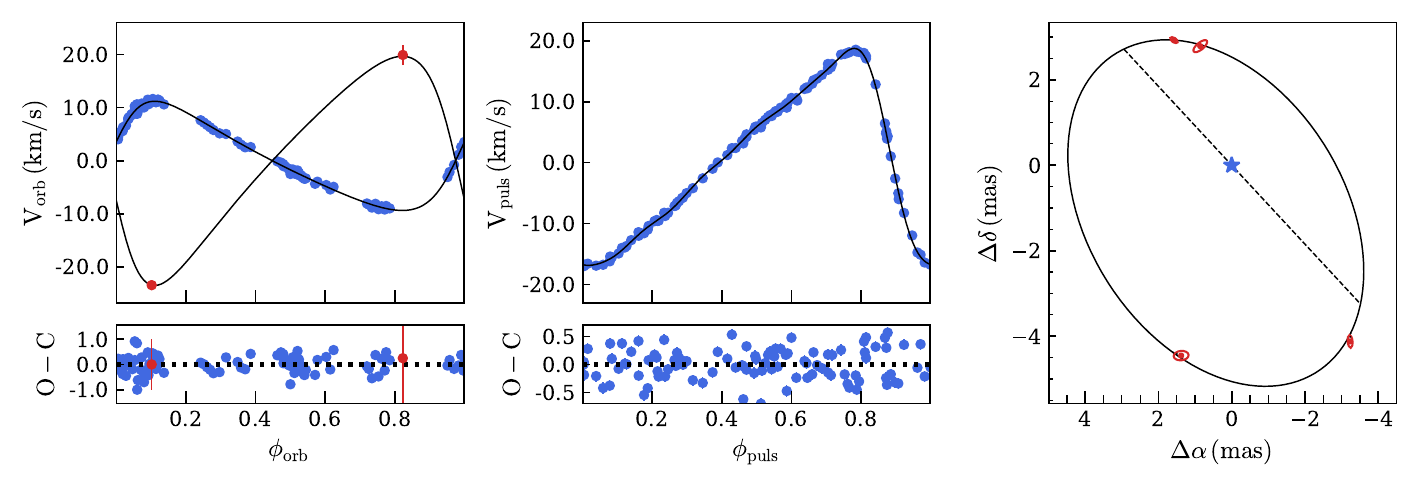}}
		\caption{Same as Fig.~\ref{figure__orbit} but including radial velocities of the companion in the left panel (red).}
		\label{figure__orbit2}
	\end{figure*}

	Evans, et al. (2018) obtained two high-resolution echelle spectra of the companion at maximum and minimum orbital velocity with the Space Telescope Imaging Spectrograph (STIS) on the HST.
        The date of first spectrum is incorrect as reported.  It should be JD 2,556,569.  However it occurs in a        flat part of the orbital velocity curve, so the orbital velocity amplitude is correct as reported.        They  were cross-correlated to obtain the orbital velocity amplitude of the companion.  These differential velocities are incorporated in our global fit with the radial velocities of the Cepheid and our measured astrometry of the companion, as detailed in Gallenne, et al. (2018b), and Gallenne, et al.  (2025). The fit including the STIS spectra is displayed in Fig.~\ref{figure__orbit2}. The measured masses of the Cepheid and the companion are $M_1 = 6.7 \pm 1.4\,$M$_\odot$ and $M_2 = 3.2 \pm 0.6\,$M$_\odot$,  and the distance to the system is $1089 \pm 85$\,pc. 
        This result for the Cepheid mass is within 1 $\sigma$ of the mass found by Evans, et al (2018).
        It is disconcerting that this value based on the most complete information about the orbit has
        the largest error, and is the most different from the value in Section~\ref{m2.orb} from
        equation 1 and M$_2$ ($M_1 = 4.7\pm0.8\,$M$_\odot$).

	A simple application of the orbital velocity ratio from Evans, et al. (2018) and M$_2$ found above,
        however, results in a Cepheid mass M$_1$ of 5.5 $\pm$ 0.4 M$_\odot$.  This is similar to the result
        from the full orbit in Section~\ref{m2.orb}.  
        The result  found by Evans, et al. (2018) from the same technique is   5.2 $\pm$ 0.3
        The companion mass
was derived from the same {\it IUE} spectrum. However the present discussion
now incorporates the Mass--Temperature relation from DEBs.  The previous
determination of the companion mass is 2.5 $\pm$ 0.1 M$_\odot$  which is
very similar to that in Table~\ref{m2}  (2.60 $\pm$  0.18  M$_\odot$)
would not account for the difference in Cepheid masses.

        \section{Summary}

   This study had two goals.  First, to derive a mass for the Cepheid V350 Sgr by including
   interferometry and the mass of the companion.  Second,  a demonstration of
   Cepheid mass determination using a combination
of the spectroscopic single lined orbit, proper motion anomalies from {\it Gaia}  and {\it Hipparcos} data,
and the companion mass from an ultraviolet spectrum.  The discussion of the different
approaches in Section~\ref{discuss} found a wide range of masses with sizable errors in general.
Presumably after the {\it Gaia} DR4 release the range of masses due to distance and orbital
inclination for systems in which it is derived from proper motion anomalies  will decrease and
allow an estimate of the source and size of mass uncertainties.  

In summary,
 the most
accurate Cepheid masses come from the combination of interferometry, spectroscopic
orbits, and orbital velocity ratios, such as SU Cyg (Gallenne, et al 2025)
which can result in masses with an accuracy of only a few per cent.
The V350 Sgr system presented here does not yet reach this accuracy.
Only a few systems are bright enough with hot enough companions for this
approach (SU Cyg and V1334 Cyg).  Interferometry for Polaris is similar
(for a cooler companion; Evans, et al 2024a).  The approach in the
present study can increase the sample considerably.

\section{Acknowledgments}

Support for JK and CP were provided from HST-GO-15861.001-A.
Institutional support was provided to NRE by the Chandra X-ray Center NASA Contract NAS8-03060.
HMG was supported through grant HST-GO-15861.005-A from the STScI under NASA contract NAS5-26555.
This project has received funding from the European Research Council (ERC) under the European Union’s
Horizon 2020 research and innovation programme  to PK (project UniverScale, grant agreement No 951549).
RIA is funded by the SNSF through an Eccellenza Professorial Fellowship, grant number PCEFP2\_194638.
This project has received funding from the European Research Council (ERC) under the European Union's Horizon 2020 research and innovation programme (Grant Agreement No. 947660). SS would like to acknowledge the Research Foundation-Flanders (grant number: 1239522N).
AG acknowledges the support of the Agencia Nacional de Investigaci\'on Científica y Desarrollo (ANID) through the FONDECYT Regular grant 1241073.
This work has made
use of data from the European Space Agency (ESA) mission Gaia (https://www.cosmos.esa.int/gaia), processed by
the Gaia Data Processing and Analysis Consortium (DPAC, https://www.cosmos.esa.int/web/gaia/dpac/consortium).
Funding for the DPAC has been provided by national institutions, in particular the institutions participating in the
Gaia Multilateral Agreement.
The SIMBAD database, and NASA’s Astrophysics Data System Bibliographic Services
were used in the preparation of this paper.

The data presented in this article were obtained from the Mikulski Archive for Space Telescopes (MAST) at the Space Telescope Science Institute. 



	\expandafter\ifx\csname natexlab\endcsname\relax\def\natexlab#1{#1}\fi
	\providecommand{\url}[1]{\href{#1}{#1}}
	\providecommand{\dodoi}[1]{doi:~\href{http://doi.org/#1}{\nolinkurl{#1}}}
	\providecommand{\doeprint}[1]{\href{http://ascl.net/#1}{\nolinkurl{http://ascl.net/#1}}}
	\providecommand{\doarXiv}[1]{\href{https://arxiv.org/abs/#1}{\nolinkurl{https://arxiv.org/abs/#1}}}

\section{Appendix A: MK Standards}

The Morgan--Keenan (1973) system of  spectral classification is a
primary tool for stellar investigations. The classes are based on
specific stars which define them.
In order to relate parameters from the ultraviolet region to 
calibrations  from other wavelength regions,
the {\it IUE} project had a program to observe
MK temperature and luminosity standards with well exposed spectra in the
low resolution mode of both the long (2000 to 3200 \AA) and short
(1180 to 1950 \AA) regions.  The results are summarized in Wu, et al. (1983,
hereafter Wu etal)
with subsequent additions (such as Wu, et al. 1991).

We have used these spectra for late B and early A main sequence stars with
our software to determine temperatures.  These temperatures can then
be compared with calibrations (including from other wavelengths).

\subsection{Observations}

The spectra used in this temperature calibration are listed in
Table~\ref{tmk}.  For the temperatures in Table~\ref{tmk}, the
results from the parabola fits are listed.  However, the errors from
the parabolas are much larger than distinctions between a good
match between spectra and atmospheres and an unacceptable match,
particularly for A stars.
Therefore the errors are take from visual inspection.

\begin{deluxetable}{lrrrrrrrr}
\tablecaption{Temperatures for MK Standard Stars\label{tmk}}
\tablewidth{0pt}
\tablehead{
  \colhead{SpTy} &  \colhead{}  & \colhead{HD}  & \colhead{E(B-V)}  & \colhead{T$_{IUE}$}  &
  \colhead{Err} &   \colhead{ Err} & \colhead{T$_{PM}$}    & \colhead{} \\ 
  \colhead{} &  \colhead{}  & \colhead{} & \colhead{mag}   & \colhead{K}  &
  \colhead{K} &   \colhead{\% } & \colhead{K}   & \colhead{} \\ 
}
\startdata
B5 V  &  $\rho$ Aur  &   34759 & 0.01  &  16800   &  400  &  2.4  & 15700    &     * \\
B6 V &   $\beta$ Sex  &  90994 & 0.00  &   14523  &  750 &      5.2 & 14500  &  \\
B8 V &  18 Tau  &   23324  &  0.04 & 13688  &  500 & 3.6  & 12500     &    * \\
B9 V &   134 Tau &   38899   &  0.00 & 11469  &  500 & 4.4 & 10700   &  \\ 
B9.5 V &  $\nu$ Cap &  193432  & 0.00  &    10672  &  250 &  2.3 &  10400   &  \\
A0 V &  $\gamma$ UMa  &  103287 & 0.01 &    9755  &  250 &  2.5 &   9700  &     * \\
A1 V &  $\delta$ UMi &   166205 &   0.00  &  9309    &  250 &  2.6 & 9200  &  \\
A2 V &  38 Lyn  &    80081  & 0.01  &   9196    &  200 &  2.2 & 8840  &  \\
A3 V &  $\alpha$ PsA  &  216956  & 0.01   & 8459      & 150 &   1.8 &  8550   &     * \\
A4 V &  $\delta$ Leo  &  97603  &  0.00  &    8243   & 150  &  1.8  &  827   &  \\
A5 V &  80 UMa &    116482   & 0.01 &  8000  &  200  &  2.5 & 8080   &  \\
\enddata
\end{deluxetable}

*MK Standard


In cooler models there is a strong feature at 1520 \AA\/ which is not
seen in the spectra.  Therefore in the fits for A0 through A4 the
region between 1517 and 1536 \AA\/ is omitted.

The spectral type is in the first column of Table~\ref{tmk}, with an
* in the final column to mark MK standards, both taken from Wu etal.
The stars are identified in columns 2 and 3.  Column 4 has E(B-V),
also from Wu etal, showing that all the stars are essentially unreddened.
The temperature from our fitting is in column 5.  Errors are listed in
columns 6 and 7, with the error from visual inspection followed by
the percentage error.  The average errors for late B stars are
approximately 4\%,
where the errors for early A stars are approximately 2\%.  Column 8
has the temperature from the calibration of Pecaut and Mamajek (2013) from
longer wavelengths.

\begin{figure}
  \plotone{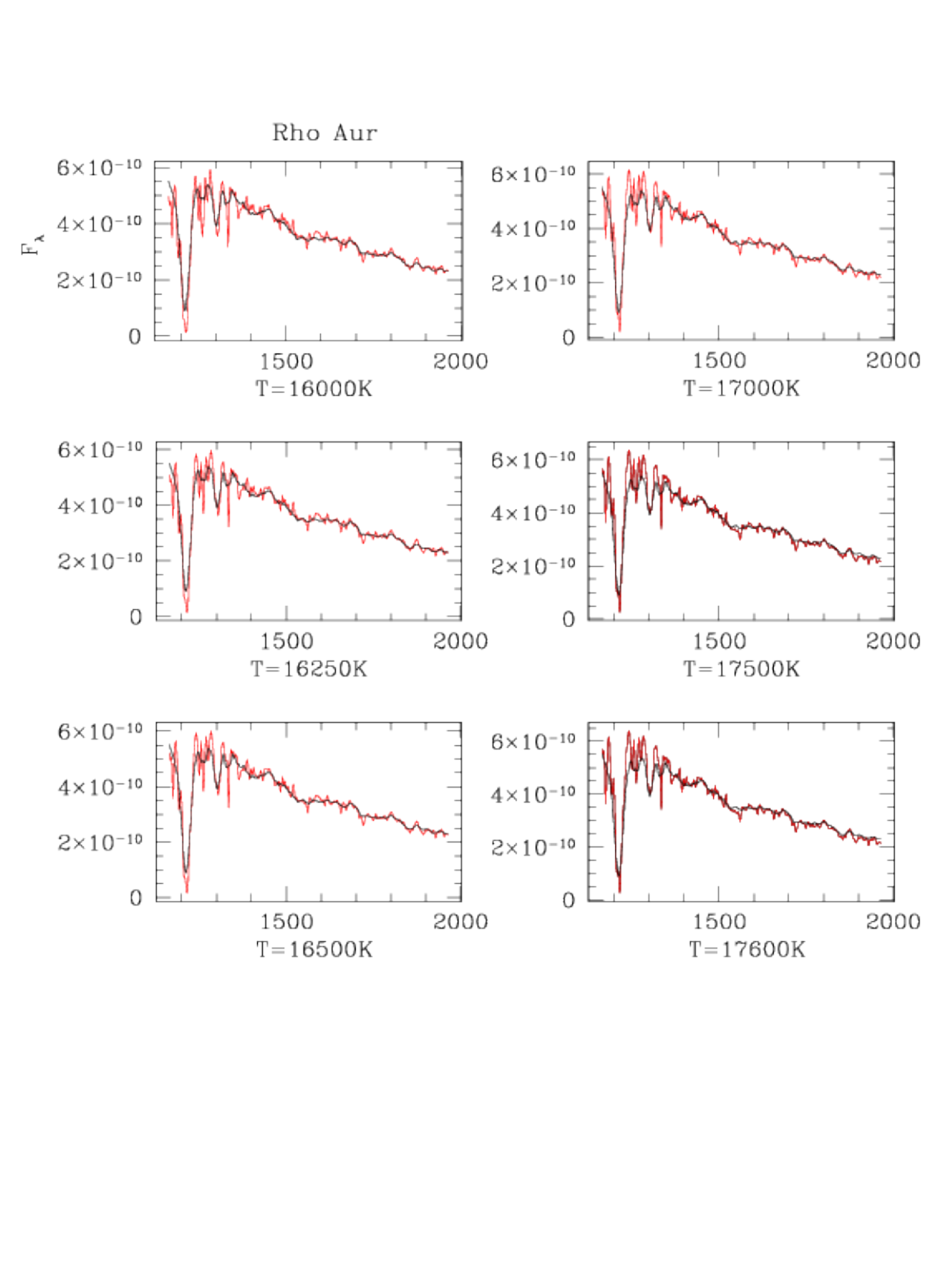}
  \caption{Energy distribution comparisons for $\rho$ Aur (B5 V).   Comparisons for the
    spectrum (black) with model atmospheres (red) for a series of temperatures from
    16000 to 17600 K. The X axis is wavelength in \AA; 
    the   Y axis is flux in ergs cm$^{-2}$ s$^{-1}$ \AA$^{-1}$.
    The  X axis is wavelength in \AA; 
 the    Y axis is flux  in ergs cm$^{-2}$ s$^{-1}$ \AA$^{-1}$. 
\label{rhoaur}}
\end{figure}

\begin{figure}
 \plottwo{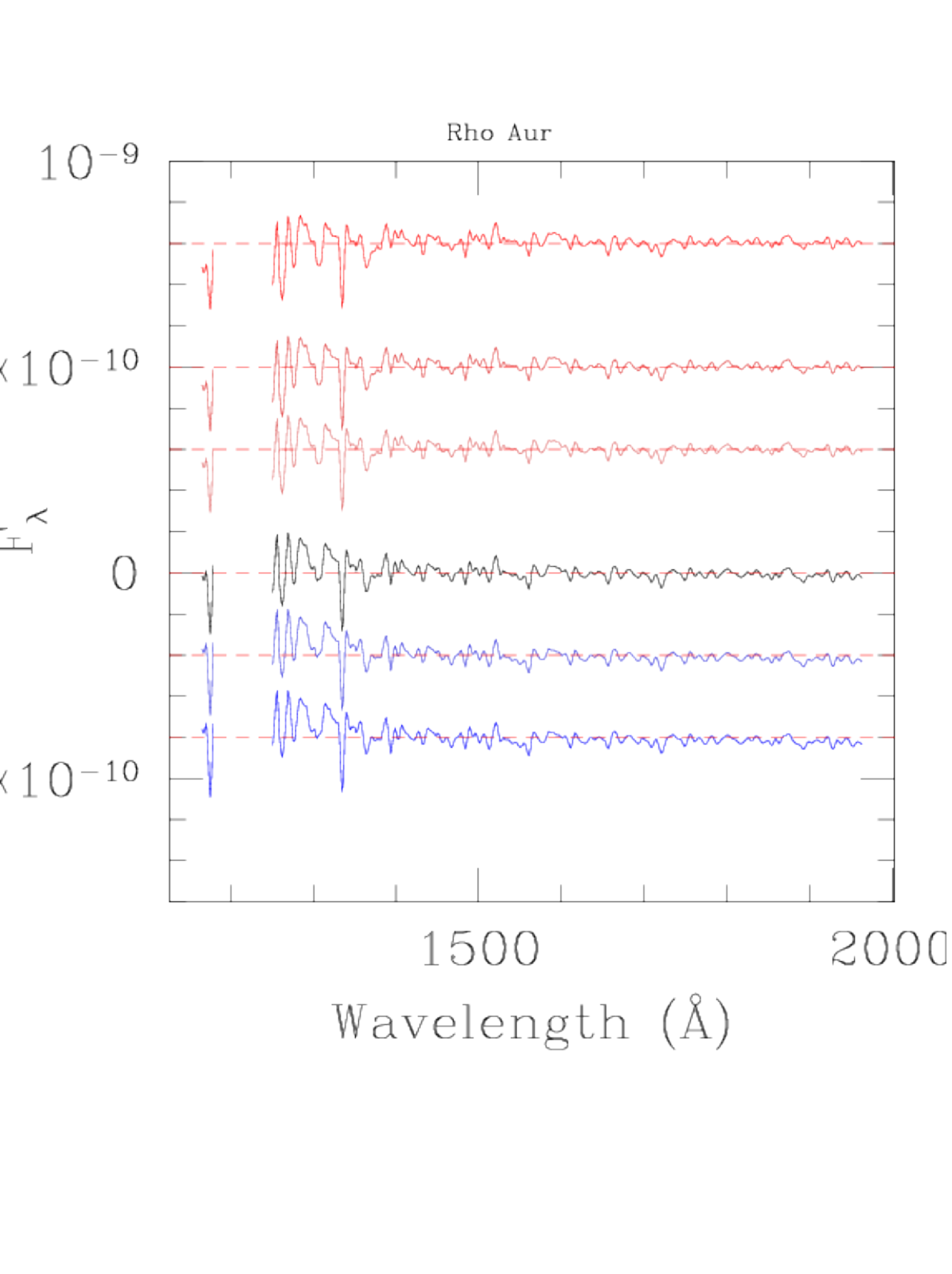}{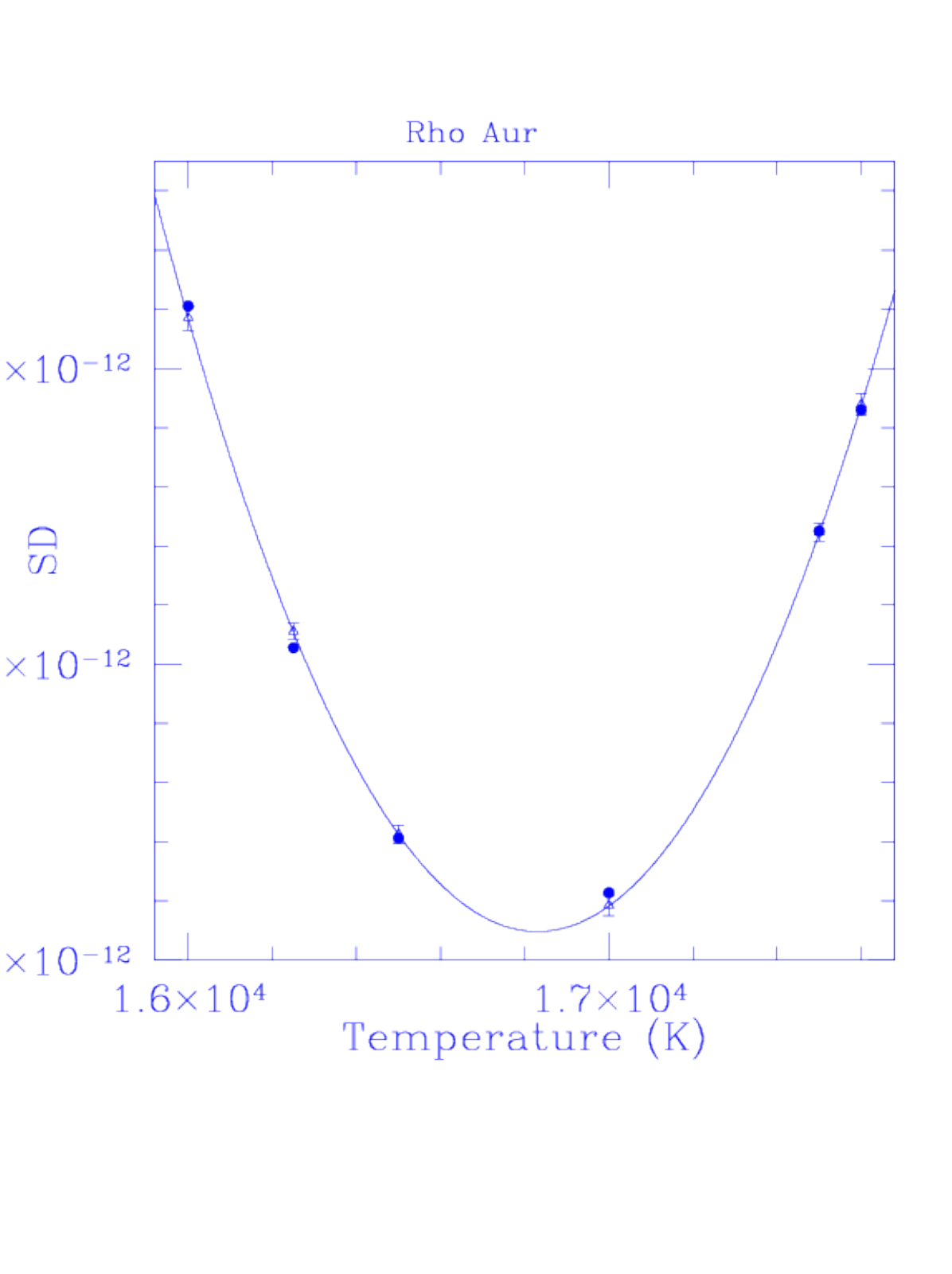}
 \caption{Energy distributions for $\rho$ Aur.
Left: The difference between the
  model  and the spectrum.  Model temperatures (starting from the top) are  16000, 16250,
   16500, 17000, 17500, and 17600 K. The  X axis is wavelength in \AA; 
 the    Y axis is flux difference in ergs cm$^2$; the models are offset for clarity.
Right:   The standard deviations from the spectrum-model comparison for
  $\rho$ Aur as the
  temperature of the models is changed.  Dots: the values of the standard deviation;
  triangles the parabola fit.
\label{rhoaurpar}}
\end{figure}

\begin{figure}
\plotone{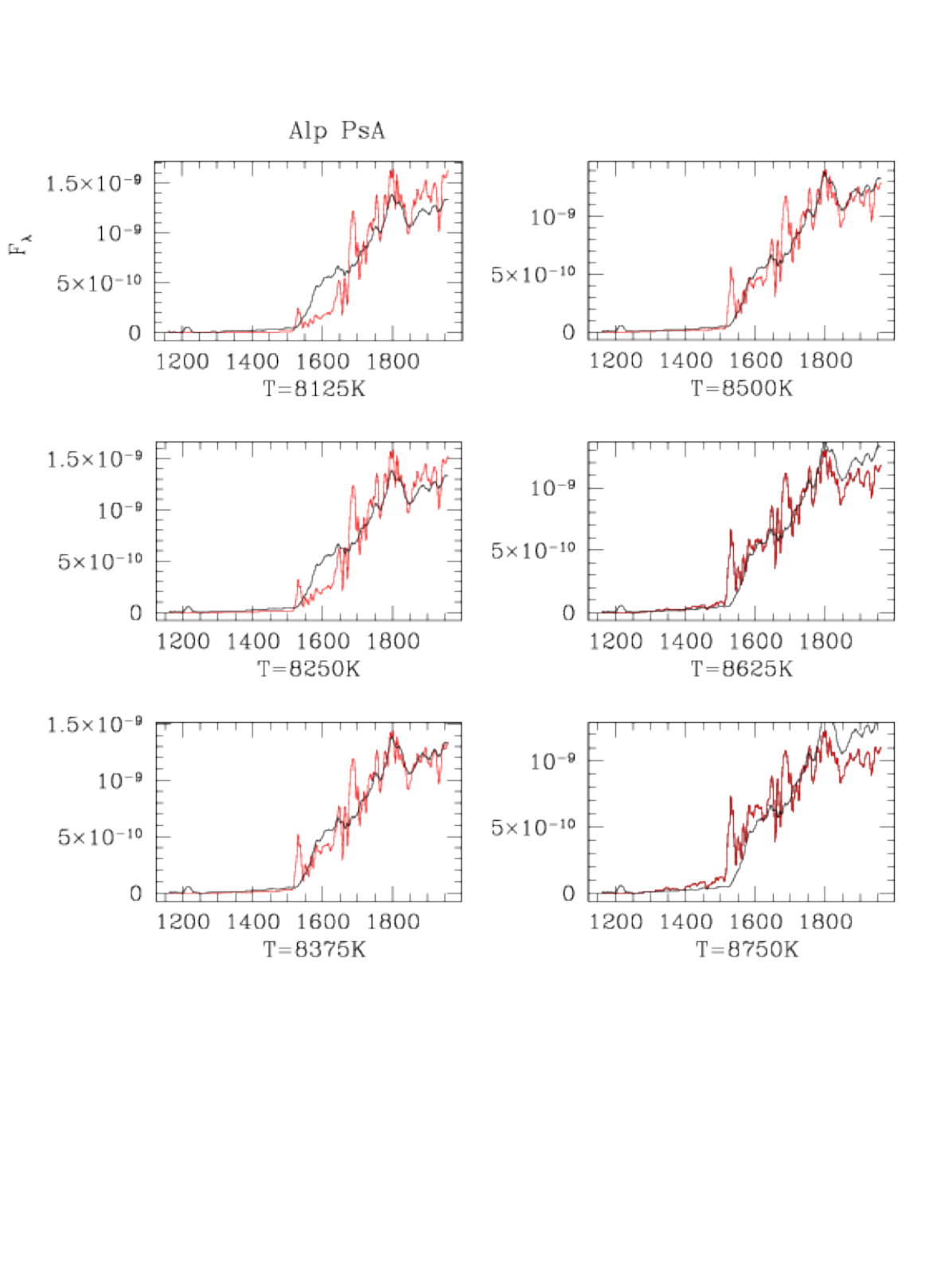}
  \caption{Energy distribution comparisons for $\alpha$ PsA (A3 V).  Comparisons for the
    spectrum (black) with model atmospheres (red) for a series of temperatures from
    8125 to 8750 K. The X axis is wavelength in \AA; 
  the  Y axis is flux in ergs cm$^{-2}$ s$^{-1}$ \AA$^{-1}$.   The X axis is wavelength in \AA; 
the  Y axis is flux  in ergs cm$^{-2}$ s$^{-1}$ \AA$^{-1}$.
\label{alppsa}}
\end{figure}

\begin{figure}
\plottwo{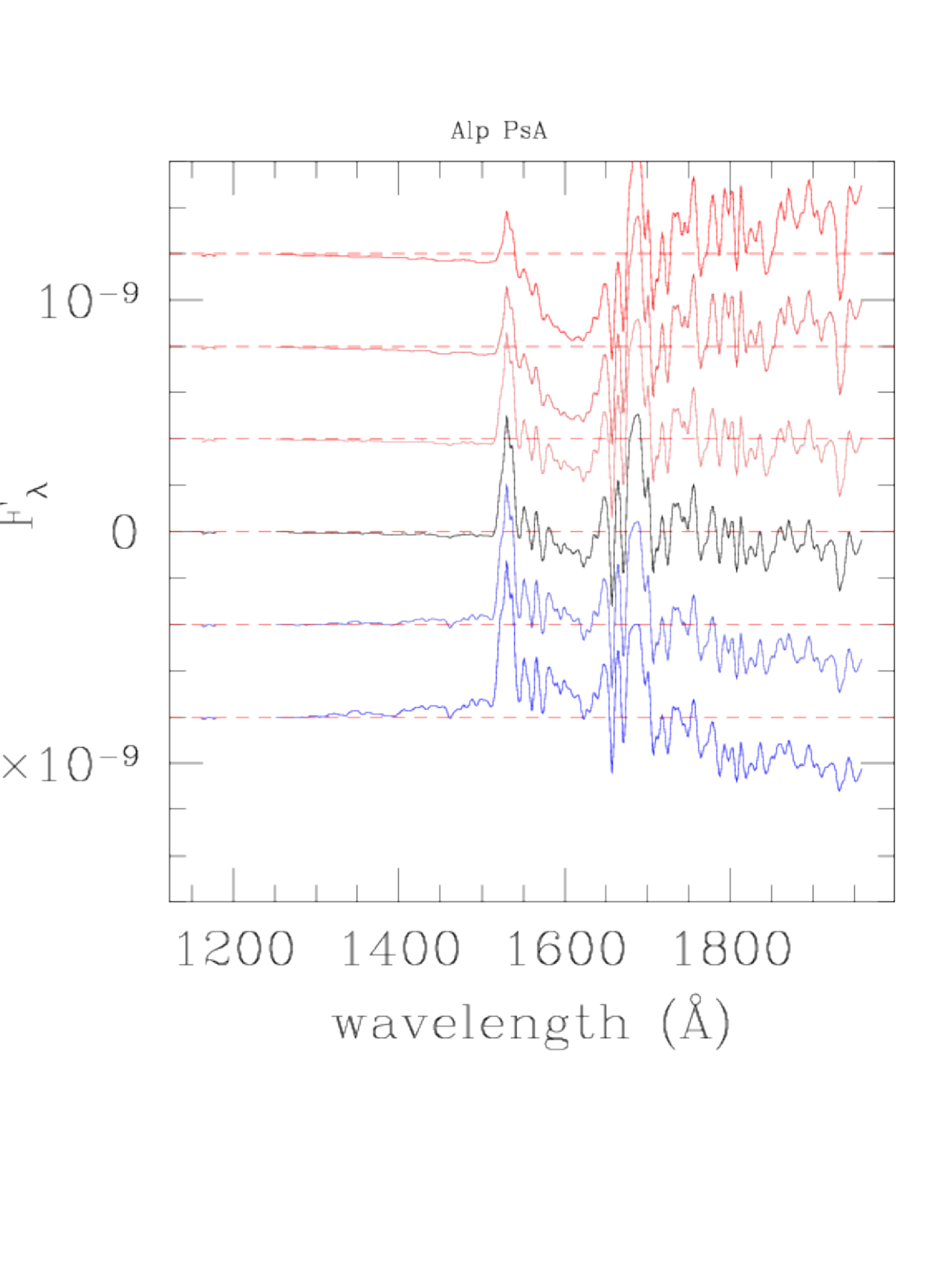}{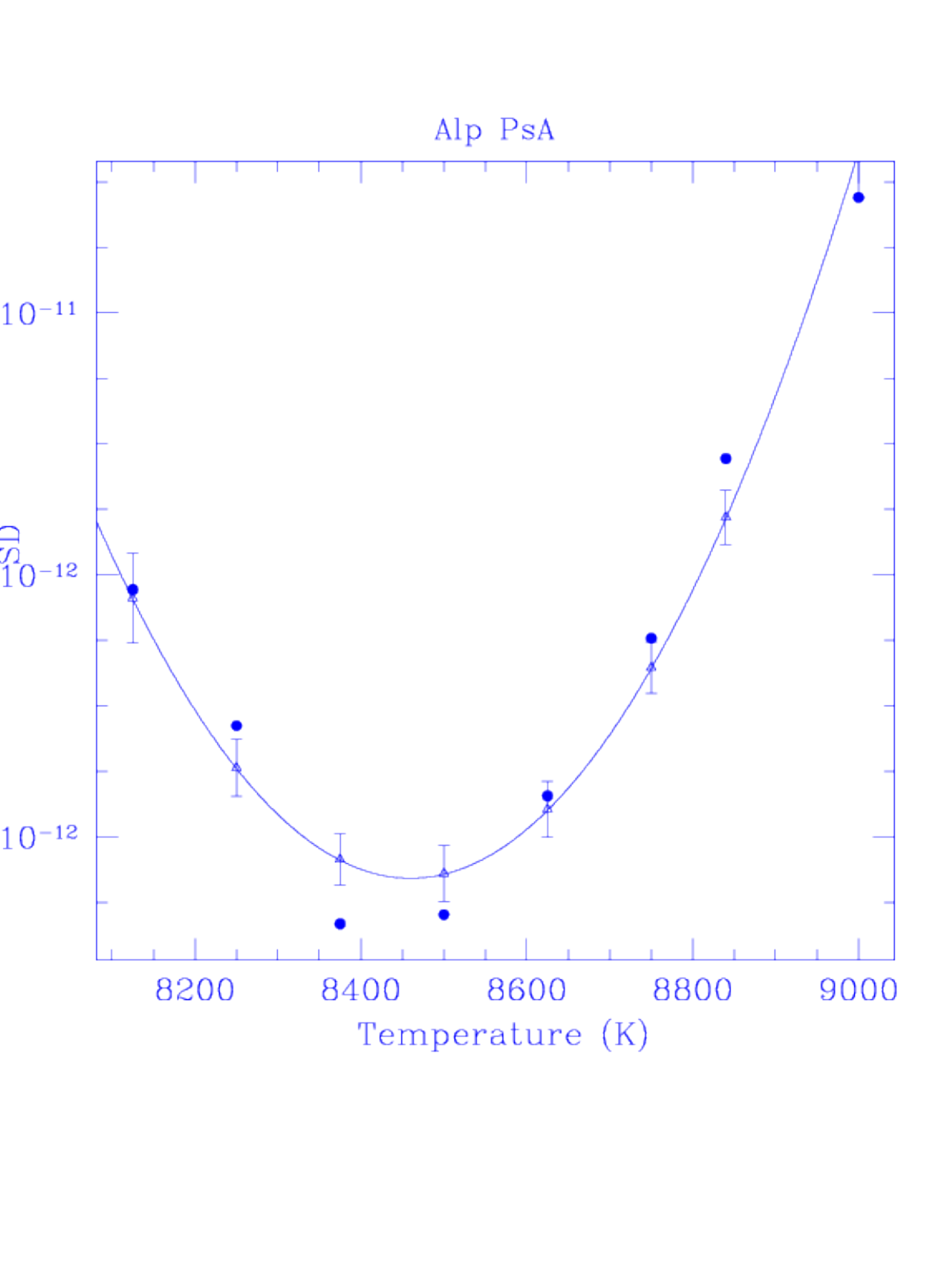}
\caption{Energy distribution comparisons for $\alpha$ PsA.  Left:
 The difference between the model
   and the spectrum.  Model temperatures (starting from the top) are  8125, 8250,
8375, 8500, 8625, and 8750    K. The models are offset for clarity.
 The X axis is wavelength in \AA; 
the  Y axis is flux difference in ergs cm$^{-2}$ s$^{-1}$ \AA$^{-1}$.
Right:  The standard deviations from the spectrum-model comparison for
  $\alpha$ PsA as the
  temperature of the models is changed.  Dots: the values of the standard deviation;
  triangles the parabola fit.
\label{alppsapar}}
\end{figure}

Examples two MK standards
of a B star ($\rho$ Aur B5 V) and an A star ($\alpha$ PsA, A3 V)
are shown in Figs~\ref{rhoaur}, ~\ref{rhoaurpar},   ~\ref{alppsa}
and  ~\ref{alppsapar}.

the models are offset for clarity.

\hfill\vfill
\eject





\subsection{Temperature Comparisons}

\begin{figure}
\plottwo{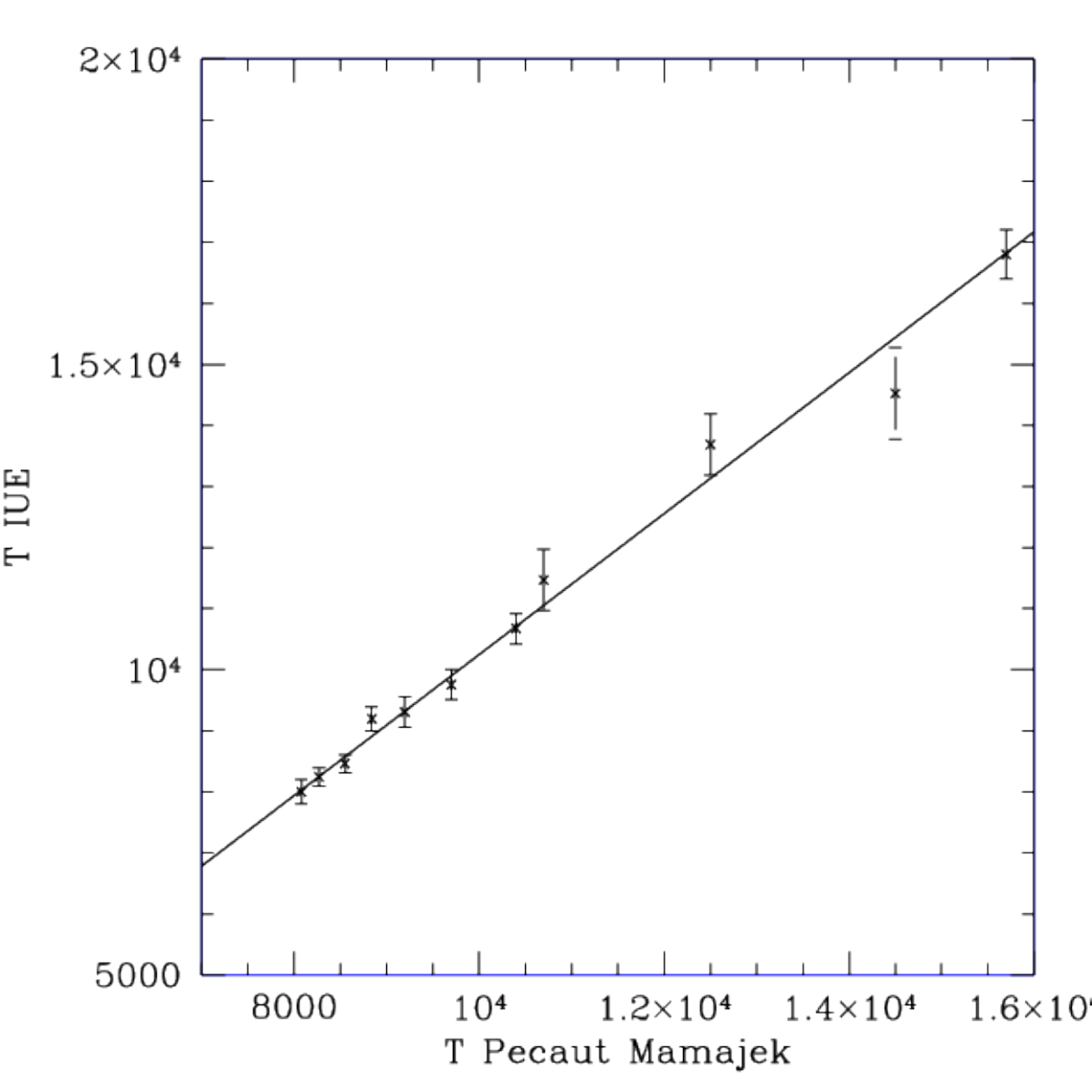}{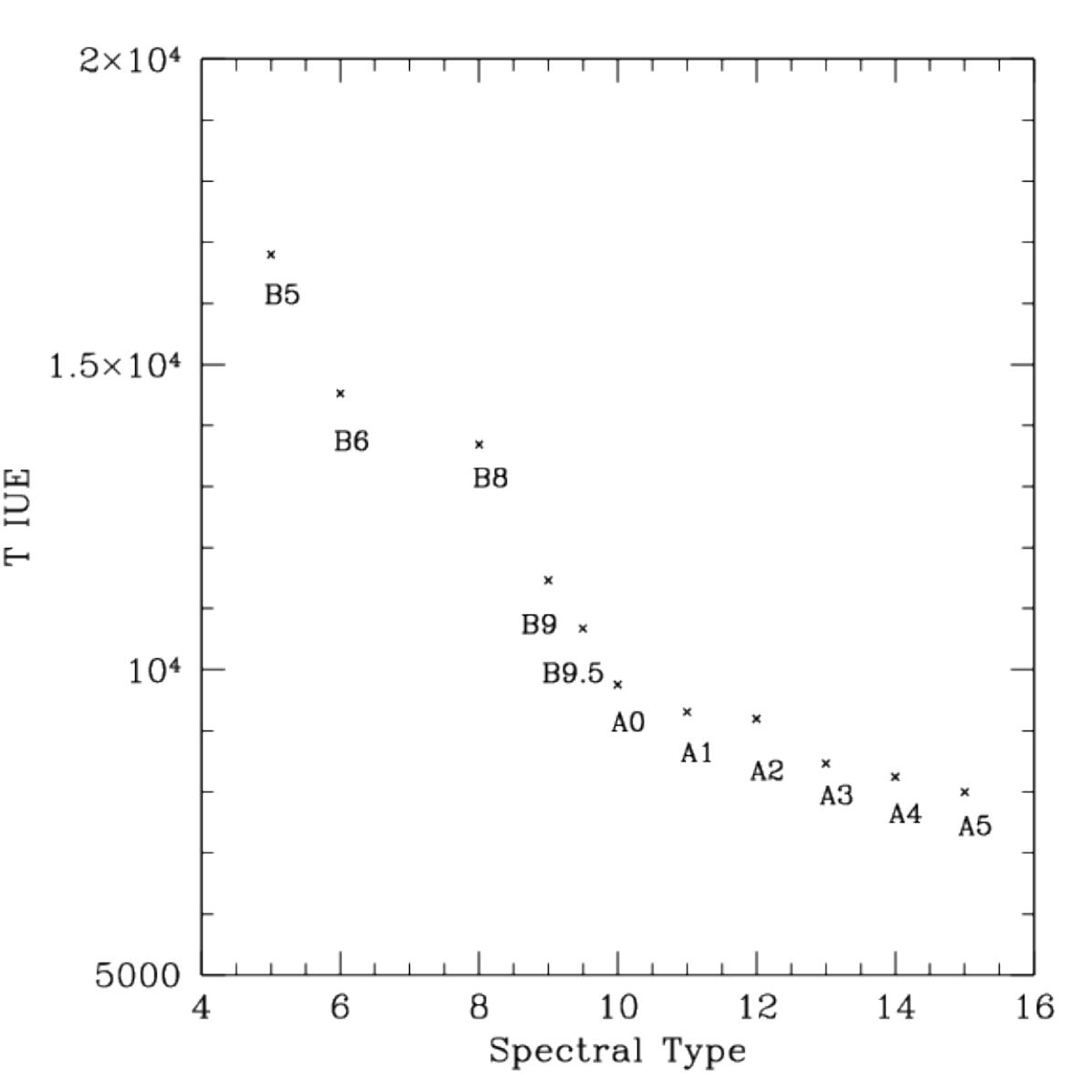}
\caption{a (Left): The temperatures from Table~\ref{tmk} from the {\it IUE} spectra
  compared with the calibration of Pecaut and Mamajek for spectral classes.
  b. (Right): T$_{IUE}$ as a function of spectral type.  The X-axis is a running number
  with 4 to 9.5 corresponding to B4 V through B9.5 V and 10 through 15
  corresponding to A0 V
  through A5 V. T is in K.
\label{tiue.pm}}
\end{figure}

Opacities in the 1200 to 1700 \AA\/ are high, and undergoing some revisions. In addition,
a comparison needs to be made of the temperatures here from the ultraviolet with
the temperature calibration for spectral classes from longer wavelengths.
Table~\ref{tmk} shows that the T$_{PM}$ are smaller than the T$_{IUE}$, as would be
expected if the opacities in the BOSZ (2017) are too large.  
In Fig~\ref{tiue.pm}a
the temperatures determined from the atmospheres in Table~\ref{tmk} are
compared with the calibration of Pecaut and Mamajek (2013).

The linear fit is shown in Fig~\ref{tiue.pm}a:

  T$_{IUE}$ =   -1298 ($\pm$ 421) +  1.154 ($\pm$ 0.045) $\times$  T$_{PM}$ 


Fig~\ref{tiue.pm}a confirms that the uncertainties from visual instection are
reasonable.

Fig~\ref{tiue.pm}b summarizes the temperatures as a function of main sequence
spectral type. Since the temperature steps between spectral type classes are
not necessarily uniform, it should be regarded as a ``finder chart''  for spectral
types.  


\end{document}